\documentclass[aps,superscriptaddress,tightenlines,nofootinbib,floatfix,longbibliography,notitlepage]{revtex4-1}
\usepackage[left=18mm,right=19mm,top=23mm,bottom=16mm]{geometry}
\usepackage{amsmath,amssymb}
\usepackage{bm}
\usepackage{comment}
\usepackage{graphicx}
\usepackage[dvipsnames]{xcolor}
\usepackage{slashed}
\usepackage{epigraph}
\usepackage[
    colorlinks=true,
    allcolors=blue
]{hyperref}
\usepackage{soul}

\usepackage{xspace}
\usepackage{bbm}

\renewcommand{\eqref}[1]{(\ref{#1})\xspace}

\let\builtinLaTeX\LaTeX
\def\LaTeX{\builtinLaTeX\xspace}

\begin{document}

\title{Superfluous Physics}

\author{Evan Berkowitz}		\email{e.berkowitz@fz-juelich.de}
							\affiliation{
	Institut f\"{u}r Kernphysik and Institute for Advanced Simulation,
    Forschungszentrum J\"{u}lich, 54245 J\"{u}lich Germany
}
\author{William Donnelly}	\affiliation{
	Perimeter Institute for Theoretical Physics, 
	31 Caroline St. N. Waterloo, ON, N2L 2Y5, Canada
}
\author{Sylvia Zhu}			\affiliation{
	Max Planck Institute for Gravitational Physics (Albert Einstein Institute),
	Callinstra{\ss}e 38, 30167 Hannover Germany
}							\affiliation{
	Leibniz Universit\"{a}t Hannover, D-30167 Hannover, Germany
}
\collaboration{Scientists Undertaking Preposterous Etymological Research Collaboration}

\date{April 1, 2019}

\begin{abstract}
	A superweapon of modern physics superscribes a wide superset of phenomena, ranging from supernumerary rainbows to superfluidity and even possible supermultiplets.
\end{abstract}

\maketitle

\section{Introduction}

\epigraph{The questions run too deep\\
For such a simple man\\
Won't you please, \\
Please tell me what we've learned?}{Supertramp, \textit{The Logical Song}}

Supermassive black holes have nonzero supermass, the quantity that couples to $\mathcal{N}=1$ supergravity, while in theories with more supercharges one may have superdupermassive black holes, and so on.
Just as the no-hair theorem tells us that non-rotating black holes can be described as massive and charged, the no-hair supertheorem tell us that non-rotating superheavy black holes can be described as supermassive and supercharged.
String theory provides hope to understand how to go beyond the semiclassical limit to describe in detail if and how black holes preserve unitarity, while superstring theory provides hope to understand the unitarity of supermassive black holes.

With the recent observation of normal black hole \cite{Abbott:2016blz,Abbott:2016nmj,Abbott:2017vtc,Abbott:2017oio,Abbott:2017gyy} and neutron star\cite{TheLIGOScientific:2017qsa} mergers via gravitational waves by the super-sensitive Advanced LIGO and VIRGO detectors, it is not unreasonable to expect gravitational-wave detections of supernovae of collapsing superstars or violent supermassive black hole astrophysical phenomena are on the horizon or, if you'll forgive the absurd pun, superhorizon. 
Indeed, the superposition of the gravitational-wave signals can be approximated as a stochastic background, just as the superimposition of all of the gravitational-wave academic publications impact the scholastic background.

The potential discovery space of astrophysical observation seems to supersede that of direct detection methods. 
Direct detection of superpartners of familiar particles, with nonzero supermass, was a potential discovery at the Superconducting Super Collider, superior to the LHC due to its souped-up center-of-mass and supercenter-of-supermass collision energy, partly due to the stupefying radius of its hoop.
When the super-expensive Superconducting Super Collider was superseded in the Congressional budget by the International Superspace Superstation, physicists' superegos were superficially bruised and a superb opportunity to supersize our superintelligence was lost.\footnote{Other supervenient inventions were no doubt lost as well.
With CERN as a guiding historical precedent, had the Superconducting Super Collider not been deemed superfluous we might have, by now, cruised our supercars down the Information Superduperhighway.}
Other direct-detection experiments such as Super-Kamiokande have successfully performed as real-time supernova monitors and detected associated neutrinos, but, unfortunately, no sneutrinos, which are expected in the Standard Supermodel.\footnote{SuperK is a Cherenkov detector consisting of about 13,000 PMTs mounted inside a superstructure of 50 kilotons, or about half a Panamax supertanker, of ultrapure water.  The PMTs are arranged in $3\times4$ arrays termed `supermodules' \cite{Fukuda:2002uc}.
SuperK also takes advantage of `Super Control Headers', a `Super-Low Energy trigger', and a `Super Memory Partner Module' to aid its search.
} We note that, although the definition of the Standard Supermodel has historically been time dependent, it has always been derived from superficial characteristics (thereby suppressing the full degree of human diversity) and generally reinforces rather than challenges the status quo \cite{fashionmodels}.

While astronomical observations of supergiants or of the supermoon are likely irrelevant, balloon experiments which monitor cosmic rays and hunt for their astrophysical origins such as SuperTIGER\cite{Binns:2014xpa} may also shed light on these phenomena.
Supertranslations of superpositions of supersymmetric supermassive black holes in minisuperspace might accrete supernova remnant winds at a super-Eddington rate via the superradiance instability; associated violent episodes should produce clear signatures.

Another superpotential avenue of discovery is super-low-energy spectroscopy.
Supersoldiering on beyond fine and hyperfine splittings, spectroscopic detections of superfine splittings may provide an opportunity for \st{ultra-} super-low energy precision experiments to snoop on the small shifts associated with otherwise virtually-unobservable virtual superpartner loops.
Another sensitive probe to supernal effects is the neutron lifetime, with recent improvements in sensitivity by UCNA, using a superratio analysis \cite{Brown:2017mhw}, and P\textsc{erkeo} III, with its supermirror guide\cite{Markisch:2018ndu}.

Henceforth, when we need to distinguish between mass and supermass we henceforth use the `super' superscript, $m^{\text{super}}$.  When referring to the supermass of a subsystem we use the `sub' subscript, $m^{\text{super}}_{\text{sub}}$.  This avoids a disastrous notation with `sub' superscripts and `super' subscripts.

\section{Superpositions}
\epigraph{I am uninterested in gravity,\\
and superuninterested in supergravity.}{Sidney Coleman\cite{vanNieuwenhuizen:2016}}

Supersymmetry has long been understood apart from its role in supergravity.
Schr\"{o}dinger, for example, presented a solution\cite{10.2307/20490744} to the quantum-mechanical Coulomb problem using operator, or supersymmetric, techniques\cite{RevModPhys.23.21}.

In normal quantum mechanics, we know that position states may be expressed as superpositions of momentum states.
In supersymmetric quantum mechanics, superposition states may be expressed as positions of supermomentum states, where the supermomentum indicates how quickly a particle moves through superspace.
Supermomentum and superposition are therefore conjugate variables, and the supermomentum operator generates supertranslations.

Bell proved normal quantum mechanics to be incompatible with local realism\cite{bell1964einstein}.
The superinduced theorem shows supersymmetric quantum mechanics to be incompatible with local superrealism.
Unfortunately, the theorem's supering does not close the superdeterminism loophole.

The race is on to build scalable quantum computers.
One strategy is to construct superconducting qubits from cuprates which, unfortunately, are not well-described by the simple Cooper pairing of BCS theory\cite{Bardeen:1957kj} or the super Cooper pairing of super Booper-Cooper-Scooper theory.
With experimental control, these systems may prove ripe for superdense coding\cite{PhysRevLett.69.2881} or pave the way towards quantum superemacy\cite{Preskill:2012tg}.

\section{Supergroups}

\epigraph{There is geometry in the humming of the strings,\\
there is music in the spacing of the spheres.}{Pythagoras}

In 1966 supersymmetry was proposed to relate mesons to baryons\cite{doi:10.1143/PTP.36.1266} without much purchase in the supermarket of ideas, but in the early 1970s was subsequently understood as a potential extension to spacetime symmetry with applications in quantum field theory and string theory \cite{Gervais:1971ji,Ramond:1971gb,Volkov:1973ix,Wess:1974tw}.
In this application, the supergroups were generated by Lie superalgebras.

Until recently a supermajority of model builders held a near-superstitious preference for supersymmetric models to cure, for example, the hierarchy problem.
While some supersymmetric models agree with all current observations\cite{fox:2005}, the lack of evidence of any low-energy supersymmetry has raised alarms, and theorists have scrambled to find new BSM theories compatible with experimental constraints.
One strategy is to invoke strong dynamics, using Yang-Mills or super-Yang-Mills.
These theories do not require any assumption about technical naturalness, as they are automatically technically supernatural\footnote{Lattice methods typically do not gauge fix and thus are, at least, ghost-free.}.

The fermions in these theories are bound by the glue, or the superglue, into baryons, which are natural dark matter candidates.
These theories may allow entire dark nuclear sectors, with stable states of more than one dark baryon.
Depending on the theory's parameters, there may be an entire phenomenology of superallowed transitions and transitions prohibited by superselection rules.
How the surviving dark matter emerged from the primordial supersoup would require the development of a dark Big Bang Nucleosynthesis.
Unfortunately, a nonperturbative description of these theories and a quantitative understanding of their properties requires lattice methods and advanced supercomputing\cite{Detmold:2014qqa,Detmold:2014kba}\footnote{Typically, superuser privileges are not needed to execute these calculations.}.
Dark matter may be responsible for periodic terrestial extinction events, including the disappearance of the supersaurs\cite{Randall:2014lxa}.

In a remarkable example of simultaneous discovery, supergroups emerged independently in 1966.
The first known example was Cream, generated by Eric Clapton, Jack Bruce, and Ginger Baker\cite{supergroups}.
Further supergroups, such as Blind Faith, Crosby, Stills, Nash \& Young, A\reflectbox{B}BA (known for such hits as ``Super Trouper''), and The Super Super Blues Band\cite{supersuperblues} were discovered and explored, largely through the late 1960s and early 1970s.
While often producing superlative superhits, many of these supergroups were transient phenomena.

\section{Supersaturation}

\epigraph{Baby, baby, baby.
}{Carlos Santana, \textit{Since Supernatural}}

While the continuous behavior of higher-order phase transitions can be characterized by supercritical exponents such as $\alpha=\text{(disappointed parent)}$, the discontinuous nature of first-order phase transitions allow for systems to be diabatically manipulated into metastable states.
One may, for example, supercool a liquid past its freezing point or superheat it past its boiling point.
Once disturbed, such a system may experience a dramatic transition to its ground state.

A chemical solution may similarly become supersaturated.
Once the solution is saturated, additional solute would typically precipitate or fail to desolve.
However, under certain circumstances, one may supersaturate a solution, increasing the solute's concentration past its saturation point.
This can be accomplished by saturating a high-temperature solvent and subsequent cooling, for example.
Just as supersaturation is more solute in less spatial volume than expected, by Lorentz invariance dissolution can also happen in less time than expected, as observed in the endochronic, or superluminal, dissolution of resublimated thiotimoline\cite{asimov:1948,asimov:1953,asimov:1960,vernon:2022}.
This rare phenomenon can likely be ascribed to the action of superoperators\cite{Deutsch:1991nm}.

The human brain may experience saturation directly.
Semantic saturation, or semantic satiation, is when a word is repeated so often that it starts to sound peculiar or lose its semantic meaning.
We speculate that there may be words or phrases that allow \emph{semantic supersaturation}, where, despite being repeated so often, past the bounds of typical semantic saturation or good taste, meaning is retained.
We know no example and a recent experiment pursued by the authors failed with a promising candidate, documented in Reference~\cite{self}.
There have also been other natural experiments which aimed for semantic supersaturation but ultimately chickened out\cite{chicken} or buffaloed the reader into confusion\cite{buffalo}\footnote{Other authors have explored similar phenomena with similar names\cite{greenberg1994bosons,goodman2015few}.}.
Nevertheless, we believe there is no fundamental obstacle to semantic supersaturation and any difficulties are ultimately superable.
It seems unlikely that experiencing semantic supersaturation requires superhuman superpowers, superheros, or, indeed, Superman\footnote{We do note, however, that Superman, being from the planet Krypton, is not human in the biological sense of the word.  We speculate that the nuclei of his supercells may carry superhelical DNA, contributing to his supersensory powers and superstrength.}\cite{tippett:2009}.

\section*{Acknowledgements}

The authors are indebted to Lydia Callis for her service during Superstorm Sandy, Shigeru Miyamoto and Takashi Tezuka for the inspirational Super Mario Bros.\cite{mario} and the Super Nintendo sequels, Masaharu Yoshii and Shinobu Toyoda for \textit{Sonic the Hedgehog 2}\cite{sonic2} and its introduction of Super Sonic, Shawn Kemp for revitalizing the Seattle SuperSonics, Jamiroquai for the song \emph{Supersonic}, and Chuck Yeager for the first supersonic flight.
No funding was provided by any SuperPAC.
We thank our supervisors for valuable advice and feedback.

\bibliography{super}

\end{document}